\documentclass[12pt]{amsart}
\usepackage{mathrsfs}
\usepackage{amssymb} %Extra symbols 
\usepackage{url}
\usepackage{hyperref}
\usepackage{xcolor}
\usepackage{graphicx} %Include postscript graphics
\usepackage{color}
\usepackage{hyperref}
\definecolor{refcolor}{rgb}{0.1,0.2,0.88}
\hypersetup{
    colorlinks,
    citecolor=refcolor,
    filecolor=refcolor,
    linkcolor=refcolor,
    urlcolor=refcolor
}

\theoremstyle{definition}
\newtheorem{theorem}{Theorem}[section]
\newtheorem{definition}[theorem]{Definition}
\newtheorem{undefinition}[theorem]{Undefinition}

\newtheorem{example}[theorem]{Example}

\newcommand{\CS}{\mathbb{S}}
\newcommand{\CT}{\mathbb{T}}

\def\({\left(}
\def\){\right)}

\newcommand{\mc}[1]{\mathcal{#1}}
\newcommand{\ms}[1]{\mathscr{#1}}

\newcommand{\op}[1]{\operatorname{#1}}

\newcommand{\R}{\mathbb{R}}
\newcommand{\N}{\mathbb{N}}

\newcommand{\de}{\op{d}}

\newcommand{\ds}{\displaystyle}

\newcommand{\ie}{\textit{i.e.}}

\newcommand{\eg}{\textit{e.g.}}

\newcommand{\sref}[1]{\S\ref{#1}}

\newcommand{\lleq}{\underline{\ll}}

\newcommand{\dsfrac}[2]{\ds{\frac{#1}{#2}}}

\newcommand{\diag}{\textnormal{diag}}

\newcommand{\vol}{\de_{vol}}

\newcommand{\image}[3]{
\begin{figure}[ht]
\begin{center}
\includegraphics[width=#2\textwidth]{#1}
\caption{\small{\label{#1}#3}}
\end{center}
\end{figure}
}

\def\hyph{-\penalty0\hskip0pt\relax}

\newcommand{\ssemiriem}{Semi{\hyph}Riemannian}
\newcommand{\semireg}{semi{\hyph}regular}

\newcommand{\nondeg}{non{\hyph}degenerate}
\newcommand{\nnondeg}{Non{\hyph}degenerate}

\newcommand{\flrw}{Friedmann-Lema\^itre-Robertson-Walker}
\newcommand{\FLRW}{FLRW}
\newcommand{\schw}{Schwarzschild}
\newcommand{\rn}{Reissner-Nordstr\"om}

\makeatletter
\renewcommand\section{\@startsection {section}{1}{\z@}%
                                   {-3.5ex \@plus -1ex \@minus -.2ex}%
                                   {2.3ex \@plus.2ex}%
                                   {\center\normalfont\Large\bfseries}}
\renewcommand\subsection{\@startsection {subsection}{2}{\z@}%
                                   {-3.5ex \@plus -1ex \@minus -.2ex}%
                                   {2.3ex \@plus.2ex}%
                                   {\normalfont\large\bfseries}}
\renewcommand\subsubsection{\@startsection {subsubsection}{3}{\z@}%
                                   {-3.5ex \@plus -1ex \@minus -.2ex}%
                                   {2.3ex \@plus.2ex}%
                                   {\normalfont\bfseries}}
\makeatother

\begin{document}

%--------------------------------------------------------
% Title
\title{Causal Structure and Spacetime Singularities}
\author{O.C. Stoica}
\date{\today. Horia Hulubei National Institute for Physics and Nuclear Engineering, Bucharest, Romania. E-mail: cristi.stoica@theory.nipne.ro, holotronix@gmail.com}

\begin{abstract}
In General Relativity the metric can be recovered from the structure of the lightcones and a measure giving the volume element. Since the causal structure seems to be simpler than the Lorentzian manifold structure, this suggests that it is more fundamental. But there are cases when seemingly healthy causal structure and measure determine a singular metric. Here it is shown that this is not a bug, but a feature, because big-bang and black hole singularities are instances of this situation. But while the metric is special at singularities, being singular, the causal structure and the measure are not special in an explicit way at singularities. Therefore, considering the causal structure more fundamental than the metric provides a more natural framework to deal with spacetime singularities.
\end{abstract}
%\pacs{04.20.Gz, 04.20.-q, 02.40.Ky, 04.60.Nc, 04.70.Bw, 98.80.-k}
\keywords{General Relativity, Causal Structure, Spacetime, Horismos, Big-Bang, Black hole, Singularities}

%--------------------------------------------------------
% Title and contents

\maketitle

%\setcounter{tocdepth}{2}
%\tableofcontents

%~~~~~~~~~~~~~~~~~~~~~~~~~~~~~~~~~~~~~~~~~~~~~~~~~~~~~~~~~~~~~~~~~~~~~~~%
\section{Introduction}
\label{s_intro}

The metric in General Relativity can be recovered from the causal or conformal structure, that is, the lightcones, up to a scaling factor. By knowing a measure defined on the spacetime, one can find the volume element, which provides the needed information to completely recover the metric, provided that the spacetime is distinguishing \cite{Zeeman1964CausalityLorentz,Zeeman1967TopologyMinkowski,Finkelstein1969SpaceTimeCode,Hawking1976NewTopologyST,Malament1977TopologySpaceTime}. Recall that a spacetime is {\em distinguishing} if its events can be distinguished by their chronological relations with the other events alone. For example, if the spacetime contains closed timelike curves, then it is not distinguishing. If spacetime is distinguishing, then the {\em horismos relation} (for two events $p,q$, we say ``$p$ horismos $q$'', and we write $p\to q$, if $q$ lies on the future lightcone of $p$) is enough to recover the causal structure \cite{Minguzzi2009HorismosGeneratesCausal}. We can even start with a reflexive relation which represents the horismos, and by imposing simple conditions, we can recover the relativistic spacetime \cite{Sto15a}.

Can we then say that {\em causal structure $+$ measure = Lorentzian spacetime}? While it is true that the metric of any distinguishing Lorentzian spacetime can be recovered from its causal structure and a measure, not any causal structure and measure lead to a Lorentzian metric. In some cases the determined metric has singularities. But this is a good thing, since we already know from the singularity theorems of Penrose and Hawking that, under very general conditions, singularities are unavoidable \cite{Pen65,Haw66i,Haw66ii,Haw67iii,Pen69,HP70}. The problem is that at singularities the metric itself becomes singular, so it seems that it is not the appropriate tool to describe singularities, and something else may be needed. I will argue in this article that the causal structure may be a better tool, and provide a better insight into singularities, than the metric.

On the other hand, the metric still can be used to describe singularities, at least in some cases. To do this, the standard mathematical framework used in General Relativity, which is {\em {\ssemiriem} Geometry}, has to be replaced with the more general {\em Singular {\ssemiriem} Geometry}, which deals with both {\nondeg} and degenerate metrics \cite{Sto11a,Sto11b}. This allowed us to rewrite Einstein's equation in a way which remains finite at some singularities where otherwise would have infinities, but which outside the singularities remains equivalent to the original equations \cite{Sto11a,Sto12b}. It also allowed us to provide finite descriptions of the big-bang singularities \cite{Sto11h,Sto12a,Sto12c}. We call these singularities, characterized by the fact that the metric becomes degenerate, bur remains smooth, {\em benign singularities}. For the black hole singularities, which usually are thought to have singularities where components of the metric tensor $g_{ab}$ tend to infinity (called {\em malign singularities}), it was shown that there are atlases where the metric is degenerate and smooth \cite{Sto11e,Sto11f,Sto11g}. This approach to singularities turned out to have an unexpected positive side effect: they are accompanied by dimensional reduction effects which were postulated in various approaches to make Quantum Gravity perturbatively renormalizable \cite{Sto12d}.

All these reasons justify the research in {\em Singular General Relativity} \cite{Sto13a}, and suggest that if we remove the constraints that the metric has to be {\em \nondeg} ({\ie} with nonvanishing determinant) everywhere, singularities turn out not to be a problem. But the metric and other geometric objects like covariant derivative and curvature seem to be different at singularities, despite the fact that the new equations treat on equal footing the singularities and the events outside them. It would be desirable to have a more homogeneous description of the spacetime, which treats even more uniformly the events at and outside the singularities.

This homogeneous description is provided by the causal structure and the measure giving the volume element. Hence, the fact that the causal structure and the measure can lead to degenerate metrics provides an extra justification of the methods of Singular {\ssemiriem} Geometry, explaining its success in the case of the benign singularities. In the same time, this suggests once more that the causal structure is more fundamental than the metric.

%~~~~~~~~~~~~~~~~~~~~~~~~~~~~~~~~~~~~~~~~~~~~~~~~~~~~~~~~~~~~~~~~~~~~~~~%
\section{Relations and intervals between events}

As long as the metric is {\nondeg}, the intervals determined by the metric between events are in correspondence with the relations determined by the causal structure. But when we start with a causal structure, the things are different, as we show in this section.

We recall first the standard definitions (see for example \cite{Penrose1972TopologyInRelativity}, then we show that they are not appropriate for singularities, and then we replace them with the appropriate ones.

%~~~~~~~~~~~~~~~~~~~~~~~~~~~~~~~~~~~~~~~~~~~~~~~~~~~~~~~~~~~~~~~~~~~~~~~%
\subsection{Standard definitions, which apply only to regular metrics}

\begin{definition}[old definition]
\label{def_old_interval}
Let $V$ be a vector space of dimension $n$, endowed with a bilinear form $g$ (called {\em metric}) of signature $(-,+,\ldots,+)$. A vector $v\in V$ is said to be
\begin{itemize}
	\item 
{\em lightlike} or {\em null} if $g(v,v)=0$,
	\item 
{\em timelike} if $g(v,v)<0$,
	\item 
{\em spacelike} if $g(v,v)>0$,
	\item 
{\em causal} if $v\neq 0$ and $g(v,v)\leq 0$.
\end{itemize}

To the vector space $V$ we associate an affine space which we will also denote by $V$ when there is no danger of confusion. The elements of the affine space $V$ are named {\em events}. Let $p\neq q$ be two distinct events from $V$, joined by a vector $v$, hence $q=p+v$.
The events $p$ and $q$ are said to be separated by a {\em lightlike}, {\em timelike} or {\em spacelike} {\em interval}, if the vector $v$ is respectively lightlike, timelike or spacelike.
\end{definition}

\image{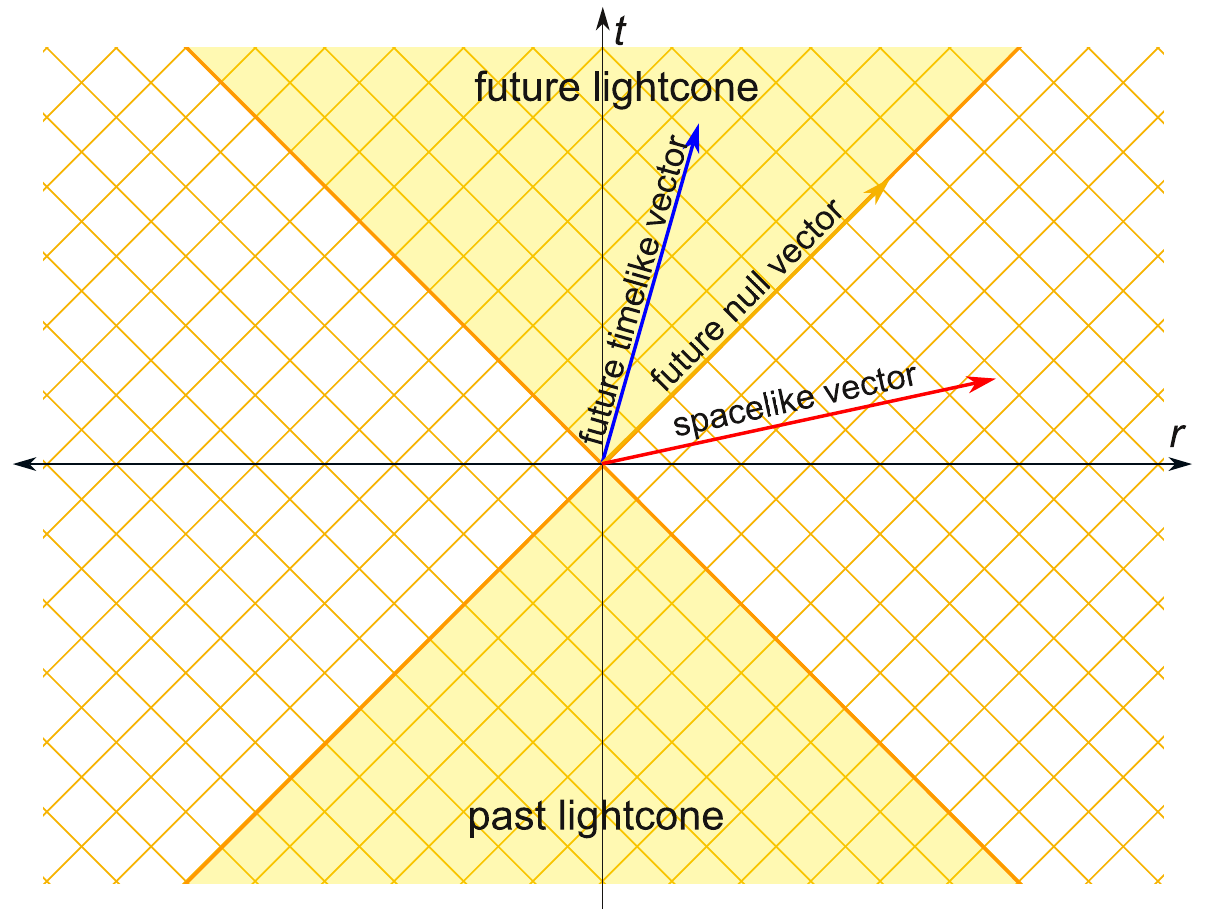}{0.65}{The causal structure of the Minkowski spacetime.}

The null vectors form the {\em lightcone}. The interior of the lightcone is made of the timelike vectors, while the exterior, on the spacelike vectors. The causal vectors form two connected components, and the choice of one of these connected components is a {\em time direction}. A causal vector from the chosen connected component is said to be {\em future-directed}, while one from the other one is called {\em past-directed} (see Fig. \ref{causal-structure-minkowski.pdf}).

With the interval between two events and the time direction one defines the following relations:

\begin{definition}[old definition]
\label{def_old_relations_minkowski}
Two events $p,q$ joined by the vector $v\in V$ are said to be in a:
\begin{itemize}
	\item 
	{\em horismos relation} $p\to q$, if $v$ is a lightlike future-directed vector,
	\item 
	{\em chronological relation} $p\ll q$, if $v$ is a timelike future-directed vector,
	\item 
	{\em causal relation} $p\prec q$, if $v$ is a causal future-directed vector,
	\item 
	{\em non-causal relation} $p\natural q$, if $v$ is a spacelike vector.
\end{itemize}
\end{definition}

These relations can be generalized to a {\em Lorentzian spacetime} $(\mc M,g)$ (a differentiable manifold $\mc M$ of dimension $n$, endowed with a metric $g$ of signature $(-,+,\ldots,+)$). First, note that the tangent space at each event $p\in \mc M$ has the structure of a Minkowski spacetime of dimension $n$, given by the metric at that point. 

\begin{definition}[old definition]
\label{def_old_curves}
Let $I\subseteq\R$ be a real interval, and $\gamma:I\to\mc M$ a curve which is differentiable everywhere. Then, the curve $\gamma$ is said to be lightlike/timelike/causal/spacelike if the tangent vectors at each of its points are lightlike/timelike/causal/spacelike. If the curve $\gamma$ is causal, it is said to be future/past-directed if the tangent vectors at each of its points are future/past-directed (with respect to its parametrization).
\end{definition}

Definition \ref{def_old_relations_minkowski} extends to any Lorentzian spacetime $\mc M$:
\begin{definition}[old definition]
\label{def_old_relations_lorentzian}
Two events $p,q\in\mc M$ are said to be in a:
\begin{itemize}
	\item 
	{\em horismos relation} $p\to q$ ($p$ {\em horismos} $q$), if they can be joined by a lightlike future curve,
	\item 
	{\em chronological relation} $p\ll q$ ($p$ {\em chronologically precedes} $q$), if $p \neq q$ and they can be joined by a timelike future curve,
	\item 
	{\em causal relation} $p\prec q$ ($p$ {\em causally precedes} $q$), if they can be joined by a causal future curve,
	\item 
	{\em non-causal relation} $p\natural q$, if $p \neq q$ and they can be joined by a spacelike curve.
\end{itemize}
\end{definition}

%~~~~~~~~~~~~~~~~~~~~~~~~~~~~~~~~~~~~~~~~~~~~~~~~~~~~~~~~~~~~~~~~~~~~~~~%
\subsection{When the metric is degenerate}

But if we start with a causal structure on a topological manifold, the standard correspondence between the intervals and causal relations no longer applies. Consider for example the causal structure of the four-dimensional Minkowski spacetime, and endow it with the metric $\tilde g=\Omega^2 g$, where $g=\diag\(-1,1,1,1\)$ and $\Omega:M\to\R$ is a smooth scalar function. As long as $\Omega\neq 0$, the causal structure determined by the metric $\tilde g$ is the same as that determined by the metric $g$ (see Fig. \ref{causal-structure-minkowski.pdf}).

\begin{example}
\label{ex_minkowski_spacelike_singularity}
Consider now that $\Omega=t$. Then, the metric $\tilde g$ is degenerate on the hyperplane $t=0$. The length of a smooth curve contained in the hyperplane $t=0$ is always vanishing, therefore even though two events $a$ and $b$ for which $t_a=t_b=0$ are not causally correlated, they can't be joined by a spacelike curve. They can be joined instead by a curve $\gamma$ so that for any event $p\in\gamma$, a vector $v_p$ tangent to $\gamma$ at $p$ satisfies $\tilde g(v_p,v_p)=0$ instead of $\tilde g(v_p,v_p)>0$. 
\end{example}

\begin{example}
\label{ex_minkowski_timelike_singularity}
Similarly, if we choose $\Omega=\sqrt{x^2+y^2+z^2}$, then the metric $\tilde g$ is degenerate on the curve $\gamma:\R\to M$, $\gamma(t)=(t,0,0,0)$. For any vector $v_p$ tangent to $\gamma$ at $p\in\gamma$, $\tilde g(v_p,v_p)=0$ instead of $\tilde g(v_p,v_p)<0$, despite the fact that any two events $a,b\in\gamma$ are in a chronological relation.
\end{example}

Examples \ref{ex_minkowski_spacelike_singularity} and \ref{ex_minkowski_timelike_singularity} show that the chronological, horismos and non-causal relations are characterized by the sign of $g_{v,v}$ only as long as the metric is {\nondeg}. If the metric is degenerate, it is possible that the relation between two events is $\ll$ or $\natural$, and yet they are not joined by a timelike or spacelike curve in the sense of the Definition \ref{def_old_curves}. This justifies that we change the definition of lightlike, timelike and spacelike curves to depend on the causal structure only, and not on the metric.

%~~~~~~~~~~~~~~~~~~~~~~~~~~~~~~~~~~~~~~~~~~~~~~~~~~~~~~~~~~~~~~~~~~~~~~~%
\subsection{General definitions}

The examples from the previous section revealed that, if we want to deal with degenerate metrics, and not only with the {\nondeg} one, we have to change the definitions of the intervals between events, and consequently of the lightlike/timelike/causal/spacelike curves, and of the relations.

\begin{undefinition}
\label{def_old_undefined}Forget Definitions \ref{def_old_interval} of intervals, \ref{def_old_curves} of causal curves, \ref{def_old_relations_minkowski} and \ref{def_old_relations_lorentzian} of relations on Minkowski and Lorentzian spacetimes.
\end{undefinition}

We will start instead from the causal structure, and consider the relations as given.
For $p\in\mc M$ we define $I^+(p):=\{r\in\mc M|p\ll r\}$, $I^-(p):=\{r\in\mc M|r\ll p\}$, $J^+(p):=\{r\in\mc M|p\prec r\}$, $J^-(p):=\{r\in\mc M|r\prec p\}$, $E^+(p):=\{r\in\mc M|p\to r\}$, and $E^-(p):=\{r\in\mc M|r\to p\}$.

The tuple $(\mc M,\to,\ll,\prec,\natural)$ is called the {\em causal structure} of the Lorentzian spacetime $(\mc M,g)$. Actually, the same information is contained in any of the triples $(\mc M,\ll,\prec)$, $(\mc M,\to,\ll)$ and $(\mc M,\to,\prec)$. If the spacetime $(\mc M,g)$ is {\em distinguishable} (that is, for any two events $p,q$ so that $I^\pm(p)= I^\pm(q)$ follows that $p=q$), then the causal structure can be recovered from $(\mc M,\to)$ alone \cite{Minguzzi2009HorismosGeneratesCausal}.

\begin{definition}[new definition]
\label{def_new_interval}
Two events $p,q$ in the Minkowski spacetime are said to be separated by a:
\begin{itemize}
	\item 
	{\em lightlike interval}, if $p\to q$ or $q\to p$,
	\item 
	{\em timelike interval}, if $p\ll q$ or $q\ll p$,
	\item 
	{\em causal interval}, if $p\prec q$ or $q\prec p$,
	\item 
	{\em spacelike interval}, if $p\natural q$.
\end{itemize}
In the Minkowski spacetime, a vector joining two events is {\em lightlike/timelike/causal/spacelike}, according to how the interval between those events is.
\end{definition}
This definition allows one to call the intervals and vectors from Examples \ref{ex_minkowski_spacelike_singularity} and \ref{ex_minkowski_timelike_singularity} spacelike, respectively timelike, despite the fact that they satisfy $\tilde g(v_p,v_p)=0$.

Now we can review Definition \ref{def_old_curves} of lightlike/timelike/causal/spacelike curves. There is no need to change it, just to plug in it the new Definition of intervals \ref{def_new_interval}, instead of the old one \ref{def_old_interval}. In fact, we can even skip altogether the differentiability of the curve (needed to discuss about tangent vectors), and characterize the curves in terms of the relations only, as we did in \cite{Sto15a}.

\begin{definition}
\label{def_causal_curve}
Let $\triangleleft$ be a relation on the events of a spacetime $\mc M$ (usually one of the relations $\to$, $\prec$, and $\lleq$, where $\lleq$ denotes ``$\ll$ or $=$'').

An {\em open curve with respect to the relation $\triangleleft$} defined on a horismotic set $\mc M$ is a set of events $\gamma\subset\mc M$ so that the following two conditions hold
\begin{enumerate}
	\item 
the relation $\triangleleft$ is {\em total} on $\gamma$, that is, for any $a,b\in \gamma$, $a\neq b$, either $a \triangleleft b$ or $b \triangleleft a$,
	\item 
for any pair $a,b\in \gamma$, $a\triangleleft b$, if there is an event $c\in\mc M\setminus\gamma$ so that $a\triangleleft c$ and $c \triangleleft b$, the restriction of the relation $\triangleleft$ to the set $\gamma\cup\{c\}$ is not total.
\end{enumerate}

We denote by $\ms C(\mc M,\triangleleft)$ the set of curves with respect to the relation $\triangleleft$.
A curve from $\ms C(\mc M,\prec)$ is called {\em causal curve}. 
A curve from $\ms C(\mc M,\lleq)$ is called {\em chronological curve}. 
A curve from $\ms C(\mc M,\to)$ is called {\em lightlike curve}. 
\end{definition}

%~~~~~~~~~~~~~~~~~~~~~~~~~~~~~~~~~~~~~~~~~~~~~~~~~~~~~~~~~~~~~~~~~~~~~~~%
\section{The causal structure of big-bang singularities}
\label{s_causal_structure_big_bang}

We will take first a look at the causal structure of the simplest big-bang cosmological model, that of {\FLRW} ({\flrw}). In this model, at any moment of time $t\in I$, where $I\subseteq\R$ is an interval, space is a three-dimensional Riemannian space $(\Sigma,g_\Sigma)$, scaled by a factor $a(t)$. The total metric is obtained by taking the {\em warped product} between the Riemannian spaces $(I,-\de t^2)$ and $(\Sigma,g_\Sigma)$, with the \textit{warping function} $a:I\to\R$,
\begin{equation}
\label{eq_flrw_metric}
\de s^2 = -\de t^2 + a^2(t)\de\Sigma^2.
\end{equation}

The typical space $\Sigma$ can be any Riemannian manifold, but usually is taken to be one that is homogeneous and isotropic, to satisfy the {\em cosmological principle}. This is satisfied by $S^3$, $\R^3$, and $H^3$, whose metric is
\begin{equation}
\label{eq_flrw_sigma_metric}
\de\Sigma^2 = \dsfrac{\de r^2}{1-k r^2} + r^2\(\de\theta^2 + \sin^2\theta\de\phi^2\),
\end{equation}
where $k=1$ for the $3$-sphere $S^3$, $k=0$ for the Euclidean space $\R^3$, and $k=-1$ for the hyperbolic space $H^3$.

The {\FLRW} solution corresponds to a fluid with mass density and pressure density represented by the scalar functions $\rho(t)$ and $p(t)$. As $a\to 0$, both $\rho(t)$ and $p(t)$ tend to infinite. But the correct densities are not the scalars $\rho(t)$ and $p(t)$, but the {\em densities} $\rho(t)\sqrt{-g}$ and $p(t)\sqrt{-g}$, which are the components of the differential $4$-forms $\rho(t)\vol$ and $p(t)\vol$. The latter are shown to remain finite in \cite{Sto11h}, because as $\rho(t)$ and $p(t)$ tend to infinite, the {\em volume form} $\vol$ tends to zero precisely to compensate them. Also, all the terms in the {\em densitized Einstein equation}
\begin{equation}
\label{eq_einstein_idx:densitized_w1}
	G_{ab}\sqrt{-g} + \Lambda g_{ab}\sqrt{-g} = \kappa T_{ab}\sqrt{-g},
\end{equation}
introduced in \cite{Sto11a}, are finite and smooth at the singularity $a=0$. This equation is equivalent to Einstein's outside the singularity, where $\det g\neq 0$. By the results presented in \cite{Sto11h,Sto12a}, we know that the {\FLRW} singularity is well behaved, despite the fact that the usual methods of {\ssemiriem} Geometry fail when the metric becomes degenerate, because instead we use the tools of Singular {\ssemiriem} geometry \cite{Sto11a}. As shown in \cite{Sto11h,Sto12a}, the solution extends naturally beyond the singularity.

Now we will show that the causal structure remains intact at the {\FLRW} singularity.

To find the null geodesics, we solve $\de s^2=0$, assuming that $a(t)=0$ when $t=0$. In coordinates $(t,r)$ the tangent of the angle made by the null geodesics and the spacelike hypersurfaces $t=0$ grows as $a$ grows, and is zero when $a(t)=0$. Hence, the null geodesics start tangential to the hypersurface $t=0$, and as the time coordinate increases, their angle grows too (see Fig. \ref{causal-structure-flrw.pdf}).

\image{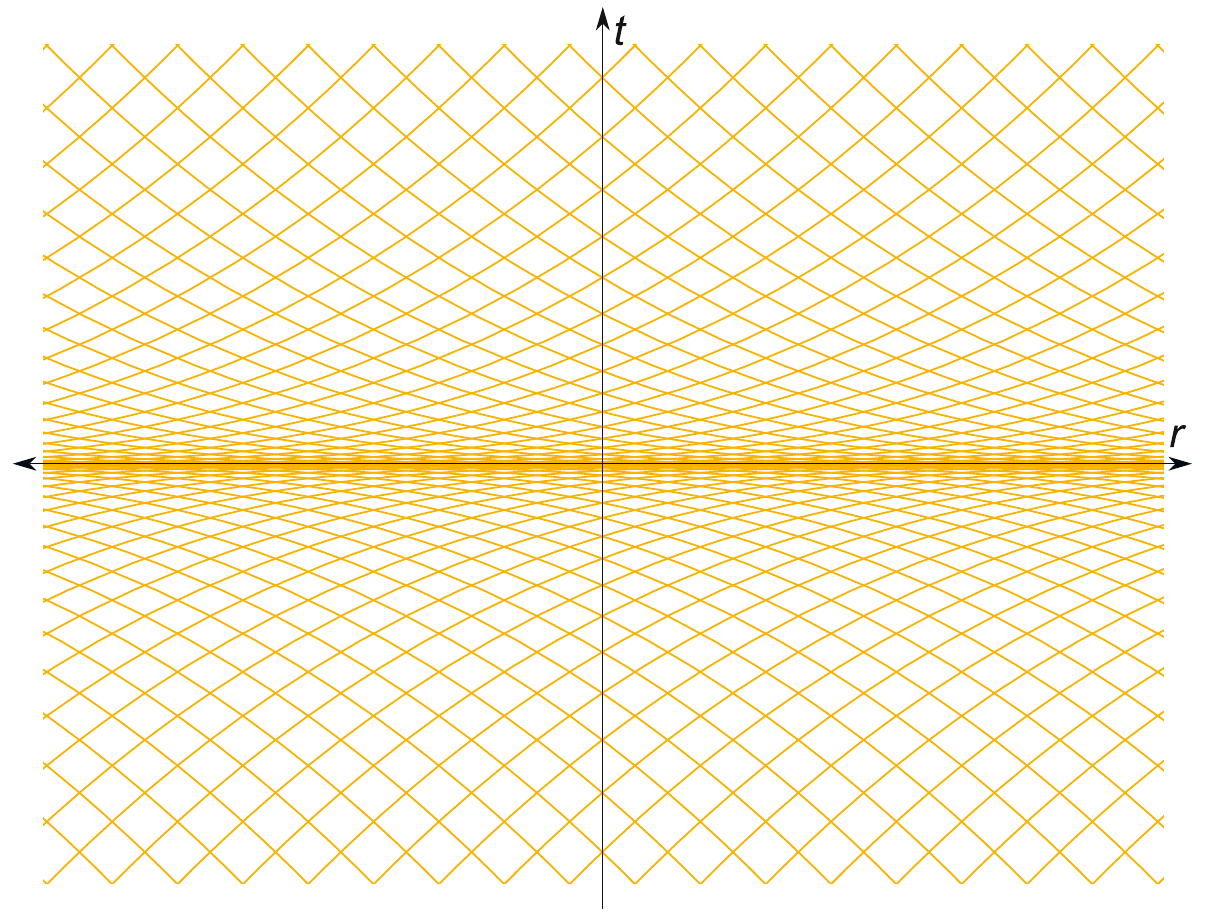}{0.65}{The causal structure of the big-bang singularity.}

The lightcones in the tangent space become degenerate at the singularity. However, the topology of a lightcone in the manifold at singularity is the same as that of one which is outside the singularity. The fact that the lightcones originating in the singularity are degenerate is a differential structure property, not a topological one.

To see this, we make the change of coordinate $t=\int a(t')\de t'$ to get $\de t=a(t')\de t'$. This puts equation \eqref{eq_flrw_metric} in the form
\begin{equation}
\de s^2 = a^2(t')\(-\de t'^2 + \de\Sigma^2\).
\end{equation}
The causal structure becomes now identical to that of the metric $-\de t'^2 + \de\Sigma^2$, which is {\nondeg}. This is to be expected, because the {\FLRW} spacetime is conformally flat. If $\Sigma=\R^3$, the causal structure becomes that of a Minkowski spacetime (Fig. \ref{causal-structure-minkowski.pdf}).

The topology of the causal structure is the same everywhere, making the causal structure universal. By contrast, the metric tensor is very different at the singularity, because it becomes degenerate.

For more generality, we can drop the conditions of homogeneity and isotropy. To do this, we allow the metric on $\Sigma$ to depend not only on time, via $a(t)$, but also on the position. So, in equation \eqref{eq_flrw_metric}, we allow $\de\Sigma^2$ to depend on time, but in such a way that it never becomes degenerate \cite{Sto12c}. The metric $\de s^2$ becomes
\begin{equation}
\label{eq_big_bang_metric}
\de s^2 = -\de t^2 + a^2(t)\de\Sigma_t^2.
\end{equation}
It is degenerate when $a=0$. We make the same change of the time coordinate, and we get that the causal structure is identical to that of the metric $-\de t'^2 + \de\Sigma_t^2$, which is {\nondeg}.

%~~~~~~~~~~~~~~~~~~~~~~~~~~~~~~~~~~~~~~~~~~~~~~~~~~~~~~~~~~~~~~~~~~~~~~~%
\section{The causal structure of black hole singularities}
\label{s_causal_structure_black_hole}

%~~~~~~~~~~~~~~~~~~~~~~~~~~~~~~~~~~~~~~~~~~~~~~~~~~~~~~~~~~~~~~~~~~~~~~~%
\subsection{The causal structure of the {\schw} singularity}
\label{s_causal_structure_schw}

The {\schw} solution represents a spacetime containing a spherically symmetric, non-rotating and uncharged black hole of mass $m$. The metric, expressed in the {\schw} coordinates, is
\begin{equation}
\label{eq_schw_schw}
\de s^2 = -\(1-\dsfrac{2m}{r}\)\de t^2 + \(1-\dsfrac{2m}{r}\)^{-1}\de r^2 + r^2\de\sigma^2,
\end{equation}
where natural units $c=1$ and $G=1$ are used. The metric
\begin{equation}
\label{eq_sphere}
\de\sigma^2 = \de\theta^2 + \sin^2\theta \de \phi^2
\end{equation}
is that of the unit sphere $S^2$ (see {\eg} \cite{HE95} p. 149).

The {\em event horizon} is at $r\to 2m$. Here apparently there is a singularity, since $g_{tt}=-\(1-\dsfrac{2m}{r}\)^{-1}\to\infty$. This singularity is due to the coordinates, and is not genuine, as one can see by using the Eddington-Finkelstein coordinates \cite{eddington1924comparison,finkelstein1958past}.

But the singularity at $r\to 0$ is genuine and malign, and can't be removed because the scalar $R_{abcd}R^{abcd}\to\infty$. However, this singularity also has a component due to the coordinates, and when we choose better coordinates, the metric becomes finite, analytic at the singularity too, even thought it still remains singular, because it becomes degenerate \cite{Sto11e}. Moreover, this degenerate singularity is of a benign, nice kind, named {\em \semireg} \cite{Sto11a}.

The new coordinates are given by 
\begin{equation}
\label{eq_coordinate_analytic}
\begin{cases}
t &= \xi\tau^4 \\
r &= \tau^2 \\
\end{cases}
\end{equation}
and the metric becomes
\begin{equation}
\label{eq_schw_semireg}
\de s^2 = -\dsfrac{4\tau^4}{2m-\tau^2}\de \tau^2 + (2m-\tau^2)\tau^4\(4\xi\de\tau + \tau\de\xi\)^2 + \tau^4\de\sigma^2,
\end{equation}
which is analytic and degenerate at $r=0$ \cite{Sto11e}.

Let's find the causal structure of this extension of the {\schw} solution at the singularity. We will consider in the following only the coordinates $(\tau,\xi)$, and the corresponding components of the metric. The full metric is obtained by taking the warped product with the metric from equation \eqref{eq_sphere}, with warping function $r(\tau)$. In coordinates $(\tau,\xi)$ the metric is analytic near the singularity $r=0$ and has the form
\begin{equation}
\label{eq_schw_metric_tau_xi_matrix}
g = (2m - \tau^2)\tau^4\left(
\begin{array}{rr}
    -\dsfrac{4}{\(2 m - \tau^2\)^2} + 16\xi^2 & 4\xi \tau \\
    4\xi\tau & \tau^{2} \\
\end{array}
\right)
\end{equation}

To find the null tangent vectors $u=(\sin\alpha,\cos\alpha)$, we have to solve for $\alpha$ the equation $g(u,u)=0$, that is
\begin{equation}
	\(-\dsfrac{4}{\(2 m - \tau^2\)^2} + 16\xi^2\)\sin^2\alpha + 8\xi\tau\sin\alpha\cos\alpha + \tau^{2}\cos^2\alpha = 0,
\end{equation}
which is quadratic in $\cot\alpha$
\begin{equation}
	\tau^{2}\cot^2\alpha + 8\xi\tau\cot\alpha + \(-\dsfrac{4}{\(2 m - \tau^2\)^2} + 16\xi^2\) = 0.
\end{equation}
The solutions are
\begin{equation}
\label{eq_null_vectors_schw}
	\cot\alpha_\pm = -\dsfrac{4\xi}{\tau} \pm \dsfrac{2}{\(2 m - \tau^2\)\tau}.
\end{equation}
Hence, the null geodesics satisfy the differential equation
\begin{equation}
\label{eq_null_geodesics_schw}
	\dsfrac{\de\xi}{\de\tau} = -\dsfrac{4\xi}{\tau} \pm \dsfrac{2}{\(2 m - \tau^2\)\tau}.
\end{equation}
They are plotted in Fig. \ref{causal-structure-schw.pdf}. We can see that the situation is very similar to that of the {\FLRW} singularity: in coordinates $(\tau,\xi)$, the null geodesics are oblique everywhere, except at $\tau=0$, where they become tangent to the hypersurface $\tau=0$.

\image{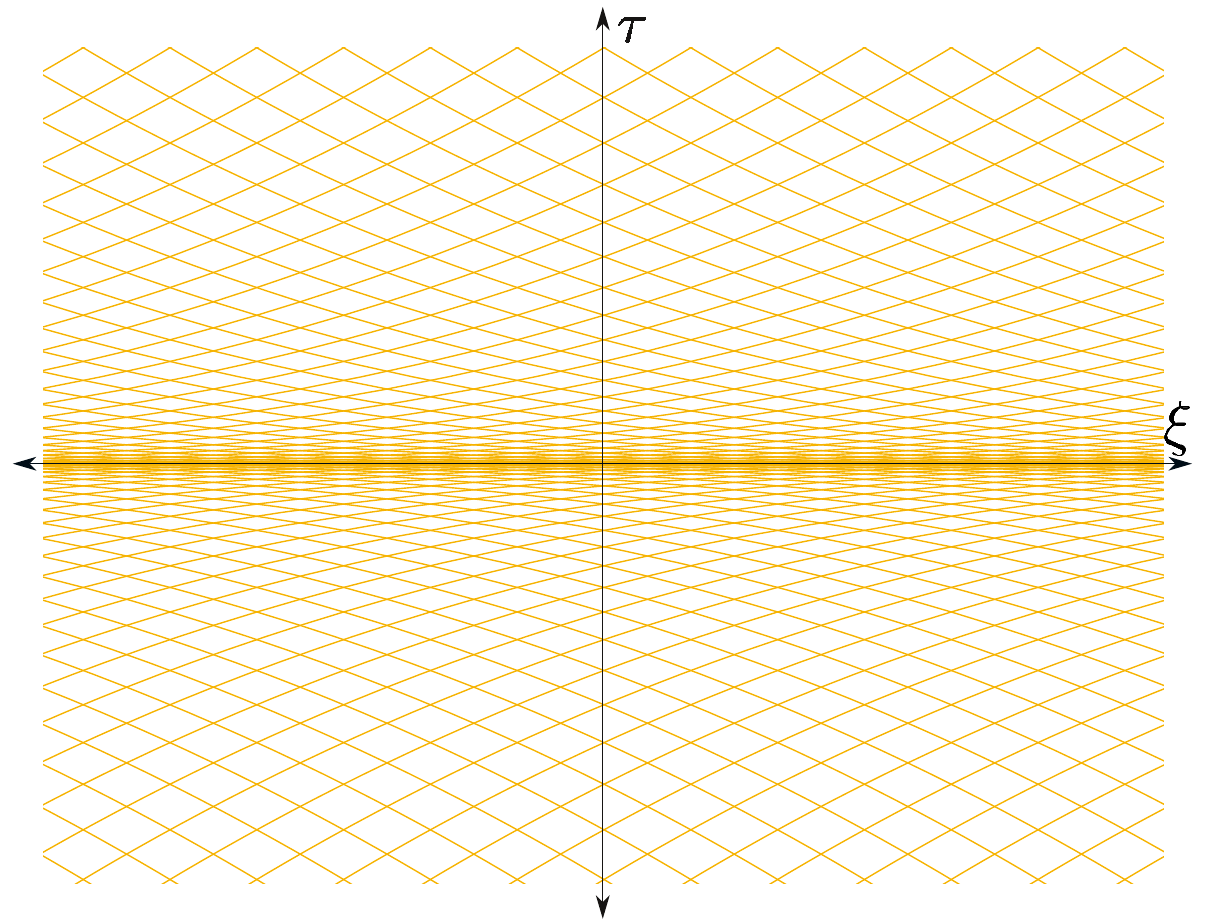}{0.65}{The null geodesics of the {\schw} solution in the $(\tau,\xi)$ coordinates.}

Since in coordinates $(\tau,\xi)$ the determinant of the metric is
\begin{equation}
\label{eq_metric_analytic_det_g_tau_xi}
\det g = - 4 \tau^{10},
\end{equation}
one may think that the metric $\tilde g=\Omega^2 g$, where $\Omega^2=\tau^{-5}$, is {\nondeg}, because its determinant is not vanishing. However, it becomes a malign singularity, since the component $\tilde g_{\tau\tau}$ becomes infinite.

However, the $4$-dimensional lightcones originating in the singularity have the same topology as any other $4$-dimensional lightcone.

%~~~~~~~~~~~~~~~~~~~~~~~~~~~~~~~~~~~~~~~~~~~~~~~~~~~~~~~~~~~~~~~~~~~~~~~%
\subsection{The causal structure of the {\rn} singularity}
\label{s_causal_structure_rn}

If the spherical non-rotating black hole of mass $m$ has an electric charge $q$, the solution is given by the {\rn} metric \cite{reiss16,nord18}, 
\begin{equation}
\label{eq_rn_metric}
\de s^2 = -\left(1-\dsfrac{2m}{r} + \dsfrac{q^2}{r^2}\right)\de t^2 + \left(1-\dsfrac{2m}{r} + \dsfrac{q^2}{r^2}\right)^{-1}\de r^2 + r^2\de\sigma^2,
\end{equation}
where $\de\sigma^2$ is that from equation \eqref{eq_sphere}, and the units are natural.

The real zeros of $\Delta:=r^2 - 2mr + q^2$ give the event horizons. The event horizons are apparent singularities, removable by Eddington-Finkelstein coordinates, just like for the {\schw} black hole. The singularity at $r=0$ can't be removed, but it can be made analytic and degenerate \cite{Sto11f}.
To do this, we change the coordinates to
\begin{equation}
\label{eq_coordinate_ext_ext}
\begin{cases}
t &= \tau\rho^\CT \\
r &= \rho^\CS, \\
\end{cases}
\end{equation}
where $\CS,\CT\in\N$.
The metric becomes 
\begin{equation}
\label{eq_rn_ext_ext}
\de s^2 = - \Delta\rho^{2\CT-2\CS-2}\left(\rho\de\tau + \CT\tau\de\rho\right)^2 + \dsfrac{\CS^2}{\Delta}\rho^{4\CS-2}\de\rho^2 + \rho^{2\CS}\de\sigma^2.
\end{equation}
The metric is analytic if
\begin{equation}
\label{eq_metric_smooth_cond}
\begin{cases}
\CS \geq 1 \\
\CT \geq \CS + 1.	
\end{cases}
\end{equation}

To find the null geodesics, we proceed as in the case of the {\schw} black hole. In coordinates $(\tau,\rho)$, the metric is
\begin{equation}
\label{eq_rn_metric_tau_rho_matrix}
g = -\Delta\rho^{2\CT-2\CS-2}\left(
\begin{array}{ll}
    \rho^2 & \CT\tau\rho \\
    \CT\tau\rho & \CT^2\tau^2 - \dsfrac{\CS^2}{\Delta^2}\rho^{6\CS-2\CT} \\
\end{array}
\right)
\end{equation}
To find the null directions $u=(\sin\alpha,\cos\alpha)$, we solve $g(u, u)=0$, which becomes
\begin{equation}
	\rho^2 \tan^2\alpha + 2\CT\tau\rho \tan\alpha + \left(\CT^2\tau^2 - \dsfrac{\CS^2}{\Delta^2}\rho^{6\CS-2\CT}\right) = 0,
\end{equation}
therefore
\begin{equation}
\label{eq_null_vectors_rn}
	\tan\alpha_\pm = -\dsfrac{\CT\tau}{\rho} \pm \dsfrac{\CS}{\Delta}\rho^{3\CS-\CT-1}.
\end{equation}
Hence, the null geodesics satisfy the differential equation
\begin{equation}
\label{eq_null_geodesics}
	\dsfrac{\de\tau}{\de\rho} = -\dsfrac{\CT\tau}{\rho} \pm \dsfrac{\CS}{\Delta}\rho^{3\CS-\CT-1}.
\end{equation}
To ensure that the coordinate $\rho$ remains spacelike, it has to satisfy $g_{\rho\rho}>0$, which is ensured in a neighborhood of $(0,0)$ by the condition $\CT \geq 3\CS$.

We see that in coordinates $(\tau,\rho)$ the null geodesics are tangent to the axis $\rho=0$, and outside the singularity $\rho=0$ they are oblique. The lightcones are stretched as approaching $\rho=0$, until they become degenerate (Fig. \ref{causal-structure-rn.pdf}).

\image{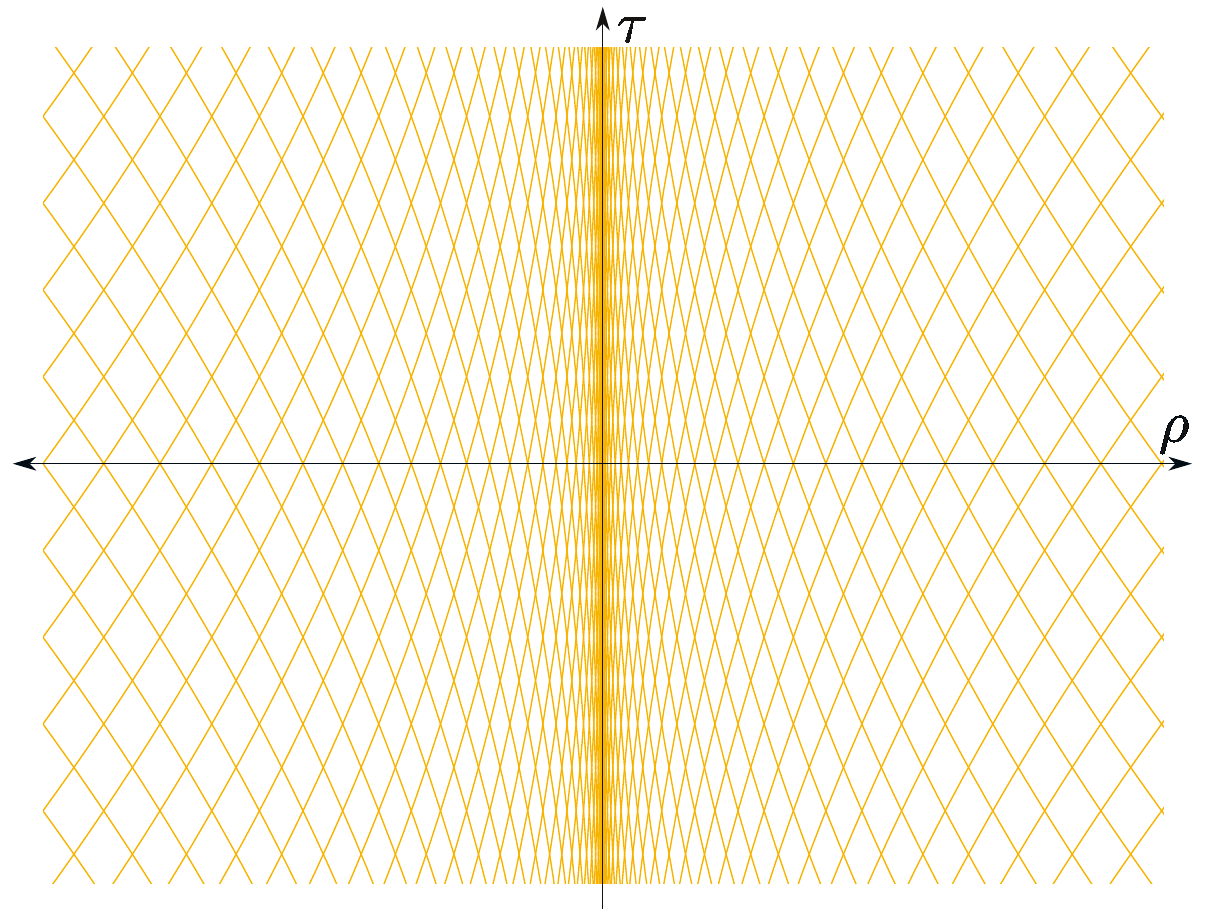}{0.65}{The null geodesics of the {\rn} solution in the $(\tau,\rho)$ coordinates, for $T\geq 3S$ and even $S$.}

%~~~~~~~~~~~~~~~~~~~~~~~~~~~~~~~~~~~~~~~~~~~~~~~~~~~~~~~~~~~~~~~~~~~~~~~%
\section{Discussion}
\label{s_discussion}

We have seen that the lightcones at two distinct events have the same topology around their origins. That is, for any two events $p$ and $q$ there are two open sets $p\in U_p$ and $q\in U_q$, and a {\em homeomorphism} (continuous bijective function) $h_{pq}:U_p\to U_q$, so that $h_{pq}(E^\pm(p)\cap U_p)=E^\pm(q)\cap U_q$. But the function $h_{pq}$ can't be always chosen to be a {\em diffeomorphism} (differentiable bijective function whose inverse is differentiable). So, lightcones are not always diffeomorphic around their origins with the other lightcones.
Figure \ref{lightcones} represents various cases of lightcones. Fig. \ref{lightcones} {\bf A}  represents a {\nondeg} lightcone, associated to a {\nondeg} metric, or to a metric that is degenerate in an isotropic manner (obtained by rescaling a {\nondeg} metric). Fig. \ref{lightcones} {\bf B} and {\bf C} represent degenerate lightcones associated to metrics degenerate in spacelike (sections \sref{s_causal_structure_big_bang}, \sref{s_causal_structure_schw}), respectively timelike directions (section \sref{s_causal_structure_rn}).

\image{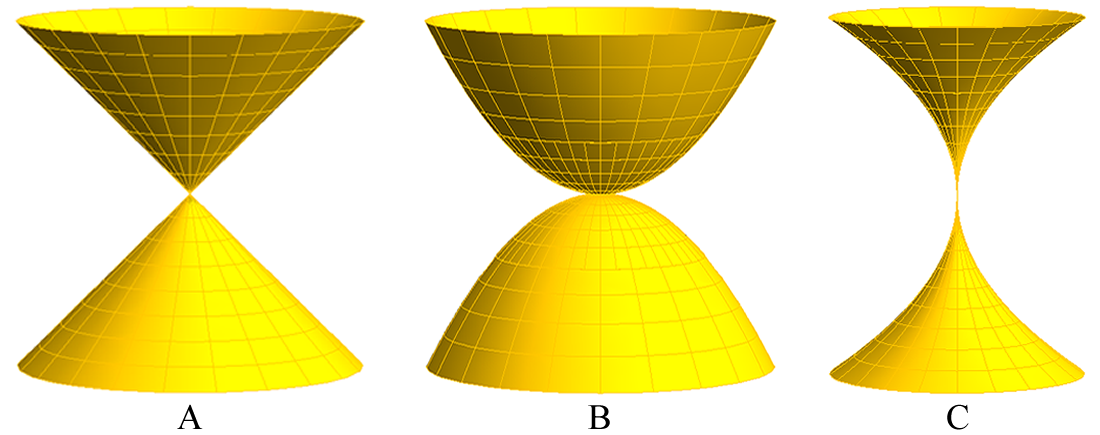}{0.8}{{\bf A--C}. Various cases of lightcones. {\bf A.} {\nnondeg} lightcone.  {\bf B.} Lightcone degenerate along spacelike directions. {\bf C.} Lightcone degenerate along timelike directions.}

%There are even cases when simply there is no {\nondeg} metric with the given causal structure, no matter how we choose the measure to determine the volume form.

The fact that lightcones are at least topologically the same around their origins allows the causal structure to be recovered from the metric not only when the metric is {\nondeg}. In the cases just described, when the metric is degenerate only along a subset $S\subset \mc M$ so that $\mc M\setminus S$ is dense in $\mc M$, the causal structure is determined at the points where the metric is {\nondeg}, and extends by continuity to the entire spacetime $\mc M$. 

The examples analyzed in the previous sections suggest that the causal structure can be seen as more fundamental, at least when the metric is allowed to become degenerate. The importance that the causal structure is maintained even at singularities can be seen from \cite{Sto12e}, where it has been shown that big-bang and black hole singularities are compatible with global hyperbolicity, which allows the time evolution of the fields in spacetime.

These results explain the success of Singular {\ssemiriem} Geometry \cite{Sto11a} and Singular General Relativity to the problem of singularities, by the fact that the causal structure is not broken at singularities, and suggests to reconstruct General Relativity starting from the causal structure.

%~~~~~~~~~~~~~~~~~~~~~~~~~~~~~~~~~~~~~~~~~~~~~~~~~~~~~~~~~~~~~~~~~~~~~~~%

\end{document}